\documentclass[aps,pre,floatfix,onecolumn,showpacs,showkeys,]{revtex4-1}
\usepackage{dcolumn}
\usepackage{bm}
\usepackage{soul}
\usepackage{color}
\usepackage{bm}
\usepackage{float}

\usepackage[colorlinks=true,urlcolor=magenta,citecolor=red,linkcolor=blue]{hyperref}

\usepackage{graphicx} \usepackage{amsmath} \usepackage{amssymb}
\newcommand{\comment}[1]{}
\newcommand{\BEQ}{\begin{equation}}
\newcommand{\EEQ}{\end{equation}}
\newcommand{\BEA}{\begin{eqnarray}}
\newcommand{\EEA}{\end{eqnarray}}

\renewcommand{\vartheta}{\theta}

\newcommand{\lar}{\leftarrow}

\begin{document}

\title{No-pumping theorem for non-Arrhenius rates}

\author{Narek H. Martirosyan}
\email{narek.h.martirosyan@gmail.com, narek@mail.yerphi.am}
\affiliation{Yerevan Physics Institute, 2 Alikhanian Brothers street, Yerevan
0036, Armenia}

\begin{abstract}
  The no-pumping theorem refers to a Markov system that holds the
  detailed balance, but is subject to a time-periodic external
  field. It states that the time-averaged probability currents nullify
  in the steady periodic (Floquet) state, provided that the Markov
  system holds the Arrhenius transition rates. This makes an analogy
  between features of steady periodic and equilibrium states, because
  in the latter situation all probability currents vanish
  explicitly. However, the assumption on the Arrhenius rates is fairly
  specific, and it need not be met in applications. Here a new
  mechanism is identified for the no-pumping theorem, which holds for
  symmetric time-periodic external fields and the so called
  destination rates. These rates are the ones that lead to the locally
  equilibrium form of the master equation, where dissipative effects
  are proportional to the difference between the actual probability
  and the equilibrium (Gibbsian) one. The mechanism also leads to
  an approximate no-pumping theorem for the Fokker-Planck rates that
  relate to the discrete-space Fokker-Planck equation.


\end{abstract}

\comment{05.70.Ln non-equilibrium processes in thermodynamics,
  thermodynamics, Fokker-Planck equation in statistical
  physics, 05.10.Gg, Fluctuation phenomena, non-equilibrium processes,
  74.40.Gh }
  

\maketitle

\section{\textbf{ Introduction.}} 

A wide range of systems appearing in physics, chemistry and biology can be modeled by Markov processes. Physically, Markov dynamics is the main tool for
describing open systems (both quantum and classical) that interact
with energy and/or particle reservoirs \cite{petru}. Hence it is at
the core of non-equilibrium thermodynamics \cite{strat}. It is also
the main tool for describing chemical reactions \cite{kampen}. Among
its biological applications one can mention conformational dynamics of
biological molecules \cite{chodera2014markov,stevens1978interactions},
ion channel gating processes\cite{sakmann2013single}, dynamics of
predation, epidemic processes, genetics of inbreeding
\cite{allen2010introduction} {\it etc}. Such applications are
frequently developed within random walk models, e.g. chemotaxis,
biological motions \cite{codling2008random} {\it etc}.

Generically, a Markov dynamics with time-independent transition rates
relaxes to a stationary state. For a single-temperature reservoir
(equilibrium thermal bath) this stationary state amounts to the Gibbs
distribution at the bath's temperature \cite{petru,kampen}. The
equilibrium nature of the bath is reflected in the detailed balance
condition that ensures nullification of all probability currents in
equilibrium \cite{petru,kampen}.

The concept of the stationary state is generalized, if the stochastic
system is subject to an external time-periodic field
\cite{petru,grim,grifoni}. The system still forgets its initial
conditions and appears in a non-equilibrium, time-dependent state,
whose probabilities oscillate with the same period as the external
field. This is the content of the Floquet theorem (outlined below),
and this motivates us to look at features of time-integrated (over
the period of the field) probability currents from one state to
another. Now the (no-pumping) theorem
\cite{rahav2008directed,chernyak2008pumping,mandal2011proof,
mandal2014unification,asban2014no}
states that probability currents nullify for an arbitrary time-periodic
external field provided that the (time-dependent) transition rate
$\rho_{i\leftarrow j}(t)>0$ from state $j$ to $i$ holds the Arrhenius
form
\begin{eqnarray}
  \label{eq:9}
  \rho_{i\leftarrow j}(t)=e^{B_{ij}+\beta E_{j}(t)},
\end{eqnarray}
where $B_{ij}=B_{ji}$ refers to the time-independent transition state,
$\beta=1/(k_{\rm B}T)$ is the inverse temperature, and $E_j(t)$ is the
oscillating energy of the state $j$. Transitions from one state to
another are induced by a thermal bath at temperature $T$, because if the bath is absent then due to energy conservation transitions between different states are also absent. Transition rates $\rho_{i\leftarrow j}(t)$ can be time-dependent solely due to an
external field that acts on the system making its energies $E_{j}(t)$
time-dependent.

Thus the no-pumping theorem shows that the non-equilibrium,
time-dependent state still holds an effectively equilibrium feature of
nullifying (time-average) currents.  (Whenever also $B_{ij}$ in
(\ref{eq:9}) are time-dependent, non-zero time-averaged currents are
not excluded.) Hence the theorem fits naturally to the continuing
effort of understanding the statistical mechanics of periodically
driven systems using analogies with the equilibrium
(i.e. time-independent) situation \cite{grifoni,kohn1,kohn2,kohn3}.
Recent works established several interesting relations between a
driven system that hold the detailed balance condition and a similar
system that is kept unnder constant (time-independent) non-equilibrium
conditions \cite{jar_r}; in this context see also
\cite{seifert_r,proesmans_r}.

Note that the same proof of the no-pumping theorem applies to rates
more general than (\ref{eq:9}), but this generalization (though useful
for its own sake) is achieved at the cost of violating the detailed
balance condition \cite{maes_r}. I.e. formally the results of
\cite{maes_r} refer to non-equilibrium baths.

The virtue of the no-pumping theorem is that it applies to all
oscillating external fields.  Its major drawback is that the Arrhenius
form (\ref{eq:9}) does not hold in many important applications, where
simultaneously the detailed balance is required. For example, the
Metropolis rates (the main tool of the Monte-Carlo dynamics), hold
(\ref{eq:9}) with $B_{ij}=-\beta\,{\rm max}[E_j(t), E_i(t)]$ (hence
$\rho_{i\leftarrow j}(t)= {\rm min}[1,e^{\beta [E_{j}(t) -
  E_{i}(t)]}]$), and $B_{ij}$ cannot stay time-independent, if
$E_i(t)$ and $E_j(t)$ are time-dependent. Further important examples
of non-Arrhenius rates include Kramers rates that emerge out of
diffusion in energy landscape \cite{kampen} and corresponds in
(\ref{eq:9}) to
\begin{eqnarray}
  \label{kramers}
\rho_{i\leftarrow j}(t)=e^{-\beta \delta_{ij} +\beta (\, E_{j}(t)-{\rm
    max}[E_i(t), E_j(t)] \,)  },
\end{eqnarray}
where $\delta_{ij} = \delta_{ji}$ is energy barrier or activation
energy that separates $E_{i}$ and $E_{j}$. Another important example is the
Fokker-Planck rates
\begin{eqnarray}
  \label{kawasaki}
\rho_{i\leftarrow j}(t)=e^{\beta[\,E_j(t)-E_i(t)  \,]/2},
\end{eqnarray}
which allow to match the discrete-space master equation for the Markov
dynamics with the continuous-space Fokker-Planck equation
\cite{agmon_h}. For all these cases, the standard formulation of the
no-pumping theorem would just allow non-zero time-averaged currents
for a suitably chosen external field, i.e. the theorem is not very
informative.

Here the aim is to extend the no-pumping theorem to rates different
from (\ref{eq:9}) (the detailed balance is always assumed to hold).

First, it will be shown that the no-pumping theorem|time-integrated
probability currents nullify|holds for the
destination rates
\begin{eqnarray}
  \label{eq:10}
  \rho_{i\leftarrow j}(t)=e^{-\beta E_{i}(t)},
\end{eqnarray}
under an additional sufficient condition that the external fields are
(effectively) time-symmetric. The mechanism is more general since it
nullifies the currents for certain non-symmetric external fields as
well.

The destination rates (\ref{eq:10}) lead to the locally-equilibrium
form of the master equation, which is driven by the difference between
the actual probability and the equilibrium one; see the discussion
after (\ref{eq:0}). There is a long and successful tradition of
applying locally-equilibrium master equations in non-equilibrium
physics. It was initiated via the model proposed in 1954 by Bhatnager,
Gross and Krook \cite{bgk,gross1,gross2}, and since that time proved
to be very useful \cite{grad}. In particular, the rates (\ref{eq:10})
were employed in \cite{de1985stretched} for describing the dynamics of
a paradigmatic disordered statistical systems (the Random Energy
Model), and found to be in agreement with experiments.  Below I show
that|in contrast to the Arrhenius rates (\ref{eq:9})|the destination
rates provide a reasonable approximation for other rates (e.g. the
Fokker-Planck rate). Hence their experimental success is not
accidental.

Second, it will be demonstrated numerically that the same mechanism
that leads to the exact no-pumping theorem for the destination rates
ensures an approximate validity of this theorem for the
Fokker-Planck rate. 

This work is organized as follows. The next section reviews the Markov
master-equations and the Floquet theorem that is necessary for
defining the no-pumping theorem. Section III is devoted to the
analytical derivation of the no-pumping theorem for the destination
rates and time-symmetric external fields.  Here I also demonstrate
that the time-symmetry is sufficient, but not necessary for the
validity of the no-pumping theorem. Section IV studies the extent to
which an approximate no-pumping theorem holds for more general rates
(e.g. the Fokker-Planck rates). Section V briefly outlines the
features of work invested to create the non-equilibrium state under
study. The results are summarized in the last section.

\section{\textbf{ Master-equation and the Floquet theorem.}} 

Consider a Markov master equation [$i,j=1,...,n,$]
\BEA
\label{eq:1}
\dot{p}_i\equiv dp_{i}/dt= {\sum}_{j}[\rho_{i\leftarrow
  j}(t)p_j-\rho_{j\leftarrow i}(t)p_i], \EEA where $p_i(t)$ is the
probability of the state $i$ at time $t$, and $\rho_{i\leftarrow
  j}(t)>0$ is the transition rate from $j$ to $i$. It is assumed that for
any fixed time $t$, there is the global detailed balance at inverse
(time-independent) temperature $\beta$:
\begin{gather}
  \label{eq:2}
  \rho_{i\lar j}(t)\, e^{-\beta E_j(t)}=  \rho_{j\lar i}(t)\, e^{-\beta E_i(t)}.
\end{gather}
Due to external field(s) acting on the system, the energies $E_i(t)$
are time-periodic functions with period $\tau$:
\begin{eqnarray}
  \label{eq:17}
  \label{fox}
E_i(t)=E_i(t+\tau).
\end{eqnarray}

The instantaneous probability flux from state $j$ to state $i$ is 
\begin{equation}
\label{bo}
J_{ij}(t) = \rho_{i\lar j}(t)p_{j}(t) - \rho_{j\lar i}(t)p_{i}(t).
\end{equation}

Before specifying the external field, let me remind the Floquet
theorem, which is necessary for defining the no-pumping theorem. Using
the normalization of probabilities $\sum_{i=1}^n {p}_i=1$, we write
(\ref{eq:1}) as 
\begin{equation}
\label{MDE}
\dot{P}=W(t)P(t)+b(t)
\end{equation}
where $P(t)=[p_1(t),...,p_{n-1}(t)]$ and $b(t)$ are $(n-1) \times 1$
vectors and $W(t)$ is $(n-1) \times (n-1)$ matrix:
\begin{gather}
b_{i} = W_{in}, ~~~ {W}_{ij} = w_{ij} - w_{in}, ~
i,j = 1, \ldots,  n-1, \\
w_{ij} = \rho_{i\lar j} - \delta_{i j}\sum_{k = 1}^{n} \rho_{k\lar
  j}, ~~ i,j = 1,~ \ldots, ~ n. 
\label{eq:4}
\end{gather}
The solution of (\ref{MDE}) with initial condition $P(t_{0})$ is 
\begin{eqnarray}
\label{ed}
{P}(t) &=& A(t,t_0) {P}(t_{0}) \nonumber\\
&+&\int_{t_{0}}^{t} {\rm d}s\,  A(t,s) b(s), ~~
A(t,s)\equiv
\overleftarrow{e}^{\int_{s}^{t}du {W}(u)},~~
\label{SOL}
\end{eqnarray}
where $\overleftarrow{e}$ is time-ordered or chronological exponent.
For $t\gg t_0$, the state ${P}(t_{0})$ is forgotten, which is
equivalent to $A(t,t_0) {P}(t_{0})\to 0$. Taking $t_0=-\infty$ in
(\ref{SOL}), and recalling that $W(t)$ and $b(t)$ are time-periodic
with the same period, we see from (\ref{SOL}) that
$P(t)=\int_{-\infty}^{t} {\rm d}s\, A(t,s) b(s)$ is also time-periodic
with the same period. This is the content of the Floquet theorem: for
sufficiently long times, the stochastic system subject to
time-periodic driving appears in a steady periodic state. This
motivates us to characterize this state via time-averaged probability
currents [cf.~(\ref{bo})]
\begin{equation}
\Phi_{ij} = \frac{1}{\tau} \int_{a}^{a + \tau} J_{ij}(t)dt,
\label{zurbagan}
\end{equation} 
where once the system is in its steady periodic (Floquet) state due to
$a \gg t_{0}$, $\Phi_{ij}$ does not anymore depend on $a$. Below I
shall employ $a=0$ in the averaging. This implies that initial
conditions are posed at much earlier time: $t_0\to -\infty$. 

Note that we do not consider cases, where the transition matrix
describes an reducible chain. There the system does not generally
forget its initial state.

\section{\textbf{ No-pumping theorem for a single field and destination
    rates.}}

Using the normalization condition for
probabilities $\sum_{j } p_{j}(t) = 1$ we obtain from (\ref{eq:1}) for
the destination rates (\ref{eq:10})
\begin{align}
\dot{p}_{i} = e^{-\beta E_{i}(t)} - p_{i}(t) Z(t), \quad
Z(t) \equiv {\sum}_{j } \, e^{-\beta E_{j}(t)}.
\label{eq:0}
\end{align}
This is the main advantage of rates (\ref{eq:10}): the equations for
the probabilities decouple from each other, making it convenient for
studying systems with irregular distribution of energies
\cite{de1985stretched}. Note that (\ref{eq:0}) can be written as
$\dot{p}_{i} =-Z(t)[\, p_{i}(t)- \frac{e^{-\beta E_{i}(t)}}{Z(t)}\,]$
showing that the change $\dot{p}_{i} $ of the probability is
proportional to the difference between this probability and its
equilibrium value $\frac{e^{-\beta E_{i}(t)}}{Z(t)}$. This makes
connection between the studied destination rates and the Bhatnager,
Gross and Krook kinetic equation \cite{bgk,gross1,gross2,grad}.


Now (\ref{eq:10}, \ref{bo}) imply
\begin{equation}
J_{ij} = \dot{p}_{i}p_{j} - \dot{p}_{j}p_{i}
= p_{j} e^{-\beta E_{i}} - p_{i} e^{-\beta E_{j}}.
\label{ord}
\end{equation}

The no-pumping statement I propose is that for 
field (\ref{fox}) [plus additional symmetry conditions to be specified
below], and for rates (\ref{eq:10}), it holds
\begin{eqnarray}
  \label{eq:3}
\langle \dot{p}_{i}p_{j}\rangle\equiv  \int_{0}^{ \tau} {\rm d}t\,
  \dot{p}_{i}(t)p_{j}(t)
  =0,
\end{eqnarray}
thereby nullifying also the time-averaged current $\Phi_{ij}=0$; see
(\ref{ord}, \ref{zurbagan}). Note
(\ref{eq:3}, \ref{eq:0}) can be written as $\langle e^{-\beta E_i}
p_{j}\rangle=\langle p_i p_{j}Z\rangle$. Hence the validity of
(\ref{eq:3}) is obvious in the limiting case 
of very slow time-dependence, where the
probabilities freeze to their Gibbsian (quasi-equilibrium) values:
$p_i(t)=e^{-\beta E_i(t)}/Z(t) $. 

To prove (\ref{eq:3}), we start from (\ref{eq:0}) and introduce there
a new time-variable $s$
\begin{eqnarray}
  \label{eq:8}
  \frac{{\rm d}s}{{\rm d}t}=Z(t), \quad s=\int_0^t {{\rm d}u}\, Z(u).
\end{eqnarray}
Due to $Z(t)>0$, the $s$-time relates to the $t$-time by a one-to-one
mapping. Since $Z(t+\tau)=Z(t)$ [see (\ref{eq:0}, \ref{fox})], we get
from (\ref{eq:8}):
\begin{eqnarray}
  \label{eq:11}
  s(t+\tau)=s(t)+\sigma, \quad \sigma= \int_0^{\tau} {{\rm d}u}\, Z(u).
\end{eqnarray}
Thus if $p_i(t)$ (in the Floquet regime) is $\tau$-periodic,
$p_i(t)=p_i(t+\tau)$, then $p_i(s)$ is $\sigma$-periodic:
\begin{eqnarray}
  \label{eq:14}
  p_i(s)=p_i(s+\sigma).
\end{eqnarray}

Note that the integral in (\ref{eq:3}) stays invariant under changing
the time:
\begin{eqnarray}
  \label{eq:13}
\int_{0}^{\tau} {\rm d}t\,
  \dot{p}_{i}(t)p_{j}(t)=\int_{0}^{\sigma} {\rm d}s\,\,
  \frac{{\rm d}p_{i}(s)}{{\rm d}s}\, p_{j}(s).
\end{eqnarray}
We get from (\ref{eq:0})
\begin{eqnarray}
  \label{eq:12}
  \frac{{\rm d}p_{i}(s)}{{\rm d}s}=
  - p_{i}(s)+e^{-\beta E_{i}(s)} /Z(s).
\end{eqnarray}

We now introduce the Fourier-expansion for $\sigma$-periodic functions 
$  g(s+\sigma)=g(s)$
\begin{eqnarray}
  \label{eq:15}
&&  g(s)=\sum_{n=-\infty}^{\infty}\hat g_n e^{\frac{2\pi {\rm i} s
      n}{\sigma}}, \\ 
&& \hat g_n=\int_0^\sigma \frac{{\rm
      d}s}{\sigma}\,
  g(s)\, e^{-\frac{2\pi {\rm i} s
      n}{\sigma}}=\int_{-\sigma}^\sigma \frac{{\rm
      d}s}{2\sigma}\,
  g(s)\, e^{-\frac{2\pi {\rm i} s
      n}{\sigma}}.~~~
  \label{baratynskii}
\end{eqnarray}
and apply it to $p_i(s)\to \hat p_{i,\, n}$ and $e^{-\beta E_{i}(s)}
/Z(s)\to \hat \psi_{i,\, n}$. Note that $\hat g_n^*=\hat g_{-n}$,
since $g(s)$ is real. Eqs.~(\ref{eq:15}, \ref{eq:12}) imply
\begin{eqnarray}
  \label{eq:20}
  \hat p_{i,\, n} = \hat \psi_{i,\, n}\left [
1+\frac{2\pi {\rm i} n}{\sigma}
\right]^{-1}.
\end{eqnarray}
Using (\ref{eq:20}) we obtain for the integral in (\ref{eq:13})
\begin{eqnarray}
\int_{0}^{\sigma} {\rm d}s\,
  \frac{{\rm d}p_{i}(s)}{{\rm d}s}\, p_{j}(s) =
  \sum_{n=-\infty}^\infty \frac{2\pi {\rm i}\,n\,\hat \psi_{i,\,
  n}\hat \psi_{j,\,
  -n}}{1+(\frac{2\pi n}{\sigma})^2}
= -\sum_{n=1}^\infty \frac{4\pi\,n\,{\rm Im}[\hat \psi_{i,\,
  n}\hat \psi_{j,\,
  -n}\,]}{1+(\frac{2\pi n}{\sigma})^2},
  \label{eq:16}
\end{eqnarray}
where we employed the Fourier expansion of (\ref{eq:12}).  If now
$\hat \psi_{i,\, n}\hat \psi_{j,\, -n}=\hat \psi_{i,\, n}\hat
\psi^*_{j,\, n}$ is real, the sum in (\ref{eq:16}) is zero.  Thus the
integrals in (\ref{eq:16}, \ref{eq:13}, \ref{eq:3}) nullify.

Let now $E_i(t)$ in (\ref{fox}) are even: 
\begin{eqnarray}
  \label{eq:7}
  E_i(t)=E_i(-t).
\end{eqnarray}
Then $s(t)$ is an odd function of $t$ [see (\ref{eq:8},\ref{eq:11})], and hence
$E_i(s)=E_i(-s)$.  Then $\hat \psi_{i,\, n}$ and $\hat \psi^*_{j,\,
  n}$ are real [see (\ref{baratynskii})], and the integral in
(\ref{eq:16}, \ref{eq:13}, \ref{eq:3}) nullifies thereby proving the
no-pumping theorem. A more general situation, when the same reasoning
applies, and $\hat \psi_{i,\, n}\hat \psi_{j,\, -n}$ is real, takes
place when $E_i(t)$ in (\ref{fox}) can be made even after a suitable
time-shift $\gamma$ which does not depend on $i$:
\begin{eqnarray}
  \label{eq:6}
E_i(t-\gamma)= E_i(-t-\gamma).
\end{eqnarray}
This is because in the Floquet regime the origin of time can be chosen
arbitrary. I stress that (\ref{eq:6}) gives only a sufficient condition for the
validity of (\ref{eq:3}). The following example illustrates this fact.
Let me take $i=1,2,3=n$ (three-level system), $\sigma=1$ and define
energies $E_i(s)$ so that the following relations hold
\begin{eqnarray}
  \label{eq:18}
e^{-\beta E_{i}(s)} /Z(s)= c_i+d_i s+f_i s^2 ~~{\rm for}~~0\leq s\leq 1,  
\end{eqnarray}
while for $s>1$ and $s<0$, $e^{-\beta E_{i}(s)} /Z(s)$ is continued
from (\ref{eq:18}) periodically with the period $\sigma=1$. These
functions are not continuous, but they can be considered as limits of
continuous functions. This suffices for the sake of the present example.



In (\ref{eq:18}), $c_i$, $d_i$ and $f_i$ are constants, which should
ensure the normalization and positivity of the probabilities
$e^{-\beta E_{i}(s)} /Z(s)$. In particular, I choose $\sum_{i=1}^3
c_i=1$ and $\sum_{i=1}^3d_i=\sum_{i=1}^3f_i=0$ for normalization. Now
generically (\ref{eq:18}) do not define any symmetric functions of
$s$. However, we get
\begin{eqnarray}
  \label{eq:19}
\hat\psi_{k,\,
    n}=\frac{f_k+{\rm i}\,(d_k+f_k)n\pi}{2n^2\pi^2}, \\
  {\rm Im}[\hat \psi_{i,\,
    n}\hat \psi_{j,\,
    -n}\,]=\frac{f_jd_i-f_id_j}{4\pi^3n^3}. 
\end{eqnarray}
The nullification of all currents amounts to ${\rm Im}[\hat \psi_{i,\,
  n}\hat \psi_{j,\, -n}\,]=0$ for all $i$ and $j$. Generally, this
requires three conditions $f_jd_i=f_id_j$ to be imposed on $f_i$ and
$d_i$. But due to $\sum_{i=1}^3d_i=\sum_{i=1}^3f_i=0$, it suffices to
take a single condition $f_1d_2= f_2d_1$. This ensures $f_jd_i=f_id_j$
and thus nullifies all currents.

\section{\textbf{Approximate no-pumping}}

\begin{figure}[H]
\centering
\includegraphics[width=10.3cm]{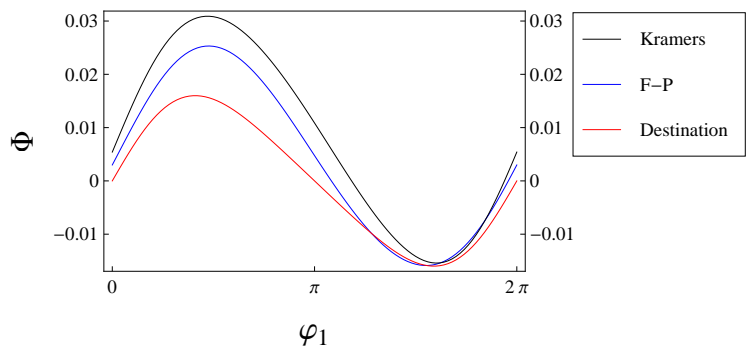} 
\caption {Time-averaged current $\Phi=\Phi_{12}=\Phi_{23}=\Phi_{31}$
  given by (\ref{zurbagan}) for a three-level system ($n=3$) and
  $\beta=1$ versus the parameter $\varphi_{1}$ for Kramers,
  Fokker-Planck (F--P) and destination rates; see (\ref{Energy}).
  $E_{i}(t)$ are given by (\ref{Energy}), where $\tau=3$. Other
  parameters in (\ref{Energy}): $\varphi_{2} = \varphi_{3} = 0$, and $
  \varepsilon_{1} = \frac{1}{3}, a_{1} = 1,\, \varepsilon_{2} =
  \frac{2}{3}, a_{2} = 2,\, \varepsilon_{3} = 1,
  a_{3} = 3 $. For Kramers rates $\delta_{ij} = 1$ in (\ref{kramers}). \\
  Hence for $\varphi_{1}=0$ or $\varphi_{1}=2\pi$, the external field
  satisfies (\ref{eq:6}) and holds the no-pumping theorem $\Phi= 0$
  for the destination rates, as seen on the figure. If (\ref{eq:6})
  holds, $\Phi\approx 0$ for the Kramers rates (\ref{eq:9},
  \ref{kramers}) and the Fokker-Planck rates (\ref{eq:9},
  \ref{kawasaki}). }
\label{fig1}
\end{figure}

\begin{figure}[htpb]
\includegraphics[width=10.3cm]{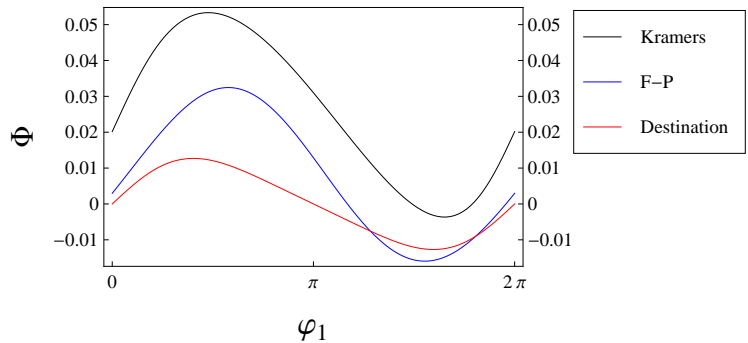} 
\caption {The same as in Fig.~\ref{fig1}, but with
  $\tau=1$, i.e. the external fields change faster than in
  Fig.~\ref{fig1}, where $\tau=3$. For this range of parameters the
  Fokker-Planck rates hold an approximate no-pumping theorem, while
  the Kramers rates do not. }
\label{fig2}
\end{figure}

\begin{figure}[htpb]
\includegraphics[width=10.3cm]{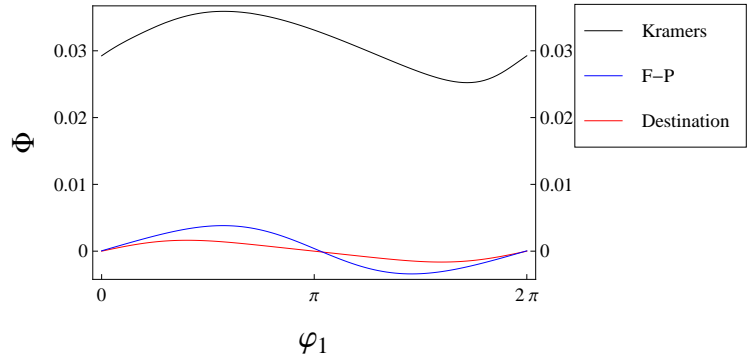} 
\caption {The same as in Fig.~\ref{fig1}, but with $\tau=0.1$;
  cf.~Fig.~\ref{fig2}. The external fields change faster than in
  Fig.~\ref{fig1} and in Fig.~\ref{fig3}. The no-pumping theorem
  approximately holds for the Fokker-Planck (FP) rates. }
\label{fig3}
\end{figure}

The above no-pumping theorem concerns the destination rates
(\ref{eq:10}). It is not valid exactly for other interesting rates,
e.g. Kramers (\ref{kramers}) or Fokker-Planck
(\ref{kawasaki}); see Figs.~\ref{fig1}--\ref{fig3}.

Now Figs.~\ref{fig1}--\ref{fig3} show numerical results, where the
time-averaged current for a three-level system is compared for three
different rates: destination (\ref{eq:10}), Kramers (\ref{kramers})
and Fokker-Planck (\ref{kawasaki}). Numerics was carried out for the
following concrete form of $E_i(t)$:
\begin{equation}
    \label{Energy}
    E_{i}(t) = \varepsilon_{i} + 
a_{i} \cos \left( \frac{2\pi t}{\tau} + \varphi_{i} \right), ~~
i=1,2,3,
\end{equation}
where $\varepsilon_{i}$, $a_{i}$ and $\varphi_{i}$ are constants.  For
(\ref{Energy}) conditions (\ref{eq:6}) are satisfied e.g. for
$\varphi_{i} = \varphi$ for all $i=1,2,3$. This situation includes,
e.g. the dipole coupling with an external, periodic
electric field \cite{petru}.

Figures \ref{fig1}--\ref{fig3} refer to different values of
the time-period $\tau$ in (\ref{Energy}).
Figs.~\ref{fig1}--\ref{fig3} demonstrate that under condition
(\ref{eq:6}), the value of the time-averaged probability current
nullifies exactly for the destination rates and it is approximately
zero (with a good precision) for the Fokker-Planck rates (denoted as
F--P in Figs.~\ref{fig1}--\ref{fig3}). For the Kramers rates the
situation is different: it also predicts an approximately zero
time-averaged probability current, but only for a sufficiently large
$\tau$; see Fig.~\ref{fig1}.



\begin{figure}[H]
\centering
\includegraphics[width=10.3cm]{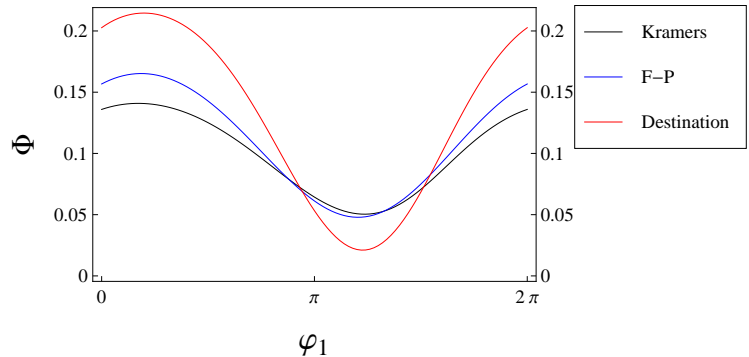} 
\caption { The same as in Fig.~\ref{fig1}, but $\varphi_{2} = \pi,
  \varphi_{3} = \frac{3\pi}{2}$. In this example conditions
  (\ref{eq:6}) do not hold and the probability currents are sizable
  for all studied rates.  }
\label{fig4}
\end{figure}

Fig.~\ref{fig4} gives an example of a situation, where (for all
studied rates) the time-averaged currents are sizable, since
conditions (\ref{eq:6}) do not hold. 

\begin{figure}[htpb]
\includegraphics[width=10.3cm]{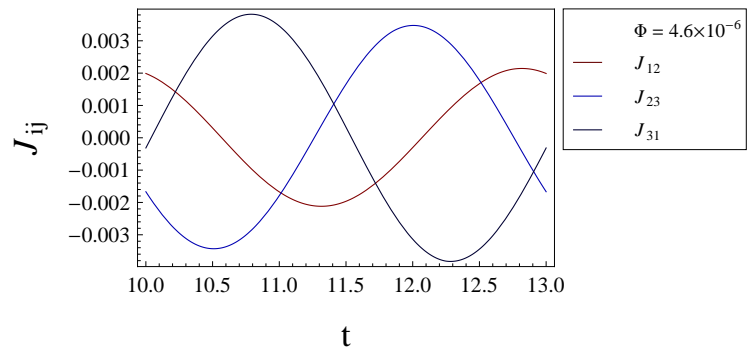} 
\caption {Instantaneous probability currents $J_{ij}(t)$ given by
  (\ref{bo}) for $\beta = 0.01$ and Kramers rates (\ref{eq:9},
  \ref{kramers}). In (\ref{Energy}) I took: $E_{i}(t) = -\frac{i}{2} +
  \frac{i}{2} \cos \left( \frac{2\pi t}{3} + \frac{\pi i}{2}
  \right) $, and for barriers:  $\delta_{ij} = 1$. \\ It is seen that $J_{ij}$ are much larger than their
  time-average $\Phi=\Phi_{12}=\Phi_{23}=\Phi_{31}$.  }
\label{fig5}
\end{figure}

Note that all above numerical examples did not refer to high
temperatures. Clearly, probability currents generally nullify for
large temperatures, but one can identify a regime, where the {\it
  instantaneous} time-dependent currents are still sizable, though
their time-averages are practically zero. This is shown in
Fig.~\ref{fig5}, where the ratio between instantaneous and averaged
currents amounts to $\sim 10^{-3}$. This high-temperature version of
the no-pumping theorem holds for all studied rates and it does not
need conditions (\ref{eq:6}). 

\begin{figure}[H]
\centering
\hspace{2em}
\includegraphics[width=10.1cm]{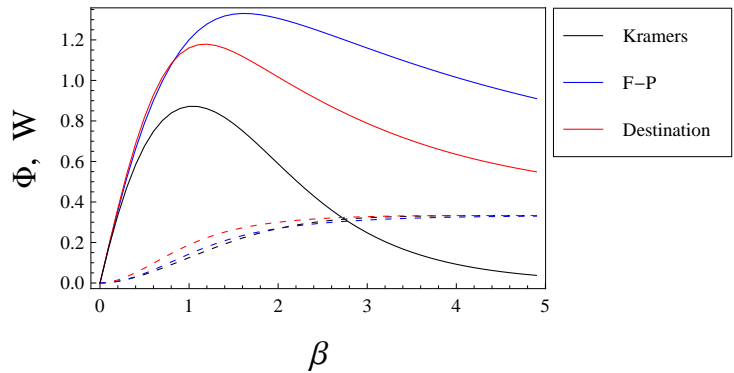} 
\caption { Time-averaged current $\Phi=\Phi_{12}=\Phi_{23}=\Phi_{31}$
  (dashed curves) given by (\ref{zurbagan}) and $W$ given by
  (\ref{eq:5}) (full curves) versus the inverse temperature $\beta$
  and for various rates. In (\ref{Energy}) I took: $E_{i}(t) =
  \frac{i}{3} + i \cos \left( \frac{2\pi t}{3} + \frac{i\pi}{2}
  \right)$.  }
\label{fig6}
\end{figure}

\section{\textbf{Work.}} 

To keep the system in the non-equilibrium state, the external field
dissipates work into the thermal bath. Now the work relates to energy
(and not probability) currents through the system. Hence it is
important to study it in the context of the no-pumping theorem.

The rate of work can
be calculated via the standard formula
\begin{equation}
\frac{{\rm d}W}{{\rm d}t}=\sum_i p_i(t)\dot{E}_i.
\end{equation}
The positivity of the time-averaged work $W$ is deduced from the
positivity of the entropy production (see e.g. \cite{seifert} for this
concept, its physical meaning is clarified below)
\begin{equation}
\label{eq:90}
S_e(t)=\frac{1}{2}\sum_{ik}(w_{ki}p_i-w_{ik}p_k)
\ln\frac{w_{ki}p_i}{w_{ik}p_k}\geq 0,
\end{equation}
where $w_{ik}$ is defined in (\ref{eq:4}). Now writing as
$\ln\frac{w_{ki}p_i}{w_{ik}p_k}=\ln\frac{w_{ki}}{w_{ik}}+\ln\frac{p_i}{p_k}$,
we note that the second term amounts in (\ref{eq:90}) to $-\frac{{\rm d
  }}{{\rm d }t}\sum_i p_i\ln p_i$ and thus disappears after the
time-averaging due to the Floquet theorem. Using (\ref{eq:2}) and
again the Floquet theorem we obtain from the first term the positivity
of the time-averaged work:
\begin{gather}
W=  \int_0^{\tau}{\rm d}t\,  \sum_i p_i(t)\dot{E}_i
  = -\int_0^{\tau}{\rm d}t\,  \sum_i \dot{p}_i(t){E}_i
=T \int_0^{\tau}{\rm d}t\,  
  S_e(t)\geq 0. 
  \label{eq:5}
\end{gather}

Applying the Clausius inequality to the bath|recall that $W$ turns to
the heat $Q$ received by the equilibrium thermal bath at temperature
$T$, and then $Q/T$ is smaller or equal to the bath entropy
increase|it is seen that $\int_0^{\tau}{\rm d}t\, S_e(t)$ gives a
lower bound for the bath entropy increase per cycle.

Note from Fig.~~\ref{fig6} that the average work decays to zero both
for high and low temperatures. There is no no-pumping
(i.e. no-work-dissipation) theorem for it.

\section{Summary}

It was shown that there is a mechanism by which the no-pumping
theorem|time-averaged probability currents nullify in the Floquet
regime|holds for the destination transition rates (\ref{eq:10}) (which
hold the detailed balance condition). A sufficient condition for this
validity is that the external time-periodic fields acting on the
stochastic system hold the time-symmetry (\ref{eq:6}). Similar
time-symmetry conditions (together with the space symmetry of the
external potential) govern the current-generation regimes of various
ratchet models; see \cite{hanggi} for a recent review. It should be
interesting to understand in more detail possible relations betweem
the no-pumping theorem and the (no) current-generation in ratchets;
this is left for future work.

In the regime, where the no-pumping theorem holds exactly for the
destination rates, there is an approximate no-pumping theorem that
holds for the physically pertinent Fokker-Planck rates.

\section*{\textbf{Acknowledgements}} I am pleased to thank
A.E. Allahverdyan for supervising this work, S.G. Babajanyan for
participating in its initial stages, and A. Yeghiazaryan for very
useful discussions. I also thank N.Z. Akopov for a careful
reading of the manuscript.

\end{document}